\documentclass[dvips,prb,twocolumn,showpacs]{revtex4}
\usepackage{amssymb}
\usepackage[dvips]{graphicx}            % to include images

\newcommand{\be}{\begin{equation}}
\newcommand{\ee}{\end{equation}}
\newcommand{\ba}{\begin{eqnarray}}
\newcommand{\ea}{\end{eqnarray}}
\newcommand{\nn}{\nonumber}

\begin{document}

\title{\textbf{Defining the effective temperature of a quantum driven
    system from current-current correlation functions}}  
\author{Alvaro Caso, Liliana Arrachea and Gustavo S. Lozano}

\address{Departamento de F\'{\i}sica, FCEyN, Universidad de Buenos Aires,
Pabell\'on 1, Ciudad Universitaria, 1428, Buenos Aires, Argentina.}

\begin{abstract}
We calculate current-current correlation functions and find an
expression for the zero-frequency noise of multiterminal systems
driven by harmonically time-dependent voltages within the Keldysh
non-equilibrium Green's functions formalism.  We also propose a
fluctuation-dissipation relation for current-current correlation
functions to define an effective temperature. We discuss the behavior
of this temperature and compare it with the local temperature
determined by a thermometer and with the effective temperature defined
from a single-particle fluctuation-dissipation relation. We show that
for low frequencies all the definitions of the temperature coincide.
\end{abstract}

\maketitle

\section{Introduction}

Over the last years we have witnessed a technological trend towards
miniaturization of electronic circuits. This tendency has been
accompanied by a growing research activity focused on achieving a
better understanding of the mechanisms for heat dissipation and energy
flow in mesoscopic systems. However, the motivation for the research
in this area is not only technological, because the very fundamental
concepts of standard statistical mechanics and thermodynamics are put
into test when studying these systems, even more when the process
under consideration corresponds to an out-of-equilibrium situation.

Several efforts have been made towards the extension of standard
thermodynamical concepts to the out-of-equilibrium evolution of
different systems. Some well-known examples are the aging regime of
glassy systems, sheared glasses, granular materials and
colloids.\cite{fdr,cukupel,letoandco} A very successful achievement in
the characterization of such nonequilibrium states has been the
identification of an {\em effective temperature}, i.e., a parameter
with the same properties of the temperature of a system at equilibrium
that is useful to describe the evolution of nonequilibrium
systems. For instance, for glassy systems the definition of effective
temperature was introduced\cite{fdr} by means of a generalization of
the equilibrium fluctuation-dissipation relations (FDR) and the
physical meaning of this concept was supported by showing that such a
temperature would coincide with the one measured by a
thermometer.\cite{cukupel} The definition of an effective temperature
from a FDR was introduced for quantum glassy systems in
Ref. \onlinecite{letogus} and later analyzed in electronic
systems\cite{lilileto}. More recently, these temperatures were studied
in an Ising chain after a sudden quench.\cite{foini}

The physics of the mesoscopic scale is ruled by the quantum coherence
of the particle propagation. This originates non-trivial interference
mechanisms and surprising effects. Well known examples are the
violation of the Fourier's Law in low-dimensional phononic
systems\cite{dhar} as well as the $2k_F$ oscillations of the local
voltage and the negative electrical resistance.\cite{fourpoint}

In the past few years there have been many experimental attempts to
locally characterize the heat flow in non-equilibrium systems. For
example, Pothier and coworkers\cite{pothier} measured the local energy
distribution function in metallic diffusive wires in a stationary
out-of-equilibrium situation. More recently, Altimiras and
coworkers\cite{altimiras} measured the electron energy distribution in
an integer quantum Hall regime with one of the edge channels driven
out-of-equilibrium. Chiral heat transport has been investigated in the
quantum Hall regime using micron-scale thermometers\cite{granger} and
later explained with the introduction of a local temperature along the
edge.\cite{fradkin} The idea of defining a non-equilibrium local
temperature has been useful to study out-of-equilibrium transport in
other mesoscopic systems. For example, thermoelectric transport has
been studied with the aid of the local temperature determined by an
ideal thermometer. \cite{casati-sanchez} Also the concept of effective
temperature has been useful to study heat exchange between a
nanojunction and its environment, which can act as a freezing agent,
\cite{Cht} and to study mesoscopic superconductors.\cite{Cht2} Another
example is the prediction that a superconducting wire can remain in
superconducting state even in contact with a bath that greatly exceeds
the critical temperature if the effective local temperature is
maintained below the critical value. \cite{dubi} A local temperature
has also been defined to characterize the heat transport in molecular
devices.\cite{galp} Additional studies have been reviewed in
Ref. \onlinecite{dubi-colloquium}.

In a previous work \cite{cal} we defined {\em local} and {\em
  effective} temperatures in electronic quantum systems driven out of
equilibrium by external ac potentials. Examples of such systems are
quantum dots with ac voltages acting at their walls (quantum
pumps)\cite{pump} and quantum capacitors.\cite{qcapexp} In that work
we presented two concepts, which are the {\em local} and the {\em
  effective} temperatures. The local temperature was introduced
following a procedure inspired in a work by Engquist and Anderson.
\cite{engq-an} The idea is to include a thermometer in the
microscopic description of the system. On the other hand the effective
temperature is defined from a local FDR involving single-particle
Green's functions. We showed that for low driving frequencies both
ways of defining the temperature coincide.  In a more recent
work\cite{cal2} we slightly generalized the definition of the
thermometer to consider the possibility of simultaneously sensing the
local temperature and the local chemical potential of the sample. We
showed that the new local temperature determined by this new
definition coincides with the previous one. Even more, we showed that
such a parameter verified the thermodynamical properties of a
temperature, meaning that its gradient signals the direction for heat
flow at the contacts.

The aim of this work is to analyze the role of effective temperatures
within the context of a FDR for current-current correlation
functions. The motivation is twofold. On one hand we are interested in
testing the robustness of the definition of an effective temperature
from a FDR, at the level of a correlation function different from the
one we have considered in our previous work. On the other hand,
current-current correlation functions are particularly appealing
quantities since they are related to noise, which can be
experimentally measured and contain valuable information on the nature
of the elementary particles that take part in the transport
process. The zero-frequency noise is usually used to characterize the
correlations between particles in mesoscopic
systems.\cite{SamButtiker} Additionally in quantum pumps, noise is
related to the possibility of having quantized pumping\cite{kamenev}
and it contains information that cannot be extracted from the
time-averaged current.\cite{ButtikerNoise0} Current correlations in
mesoscopic coherent conductors were first discussed by M. B\"uttiker
in Ref. \onlinecite{buttiker90} and since then an extensive
theoretical literature on noise in mesoscopic systems analyzed within
the scattering matrix formalism has been
developed. \cite{reviews,ButtikerNoise0,ButtikerNoise,hanggi} We use
here another approach, which is based Keldysh formalism. For
non-interacting systems both treatments were proved to coincide at the
level of the description of the current for dc\cite{fisher-lee} and
ac-driven systems.\cite{liliflo} In the present work we show that this
is also the case for the current fluctuations correlations.  The main
goal of this work is to show that the effective temperature obtained
from a fluctuation-dissipation relation for current-current
correlation functions coincides with the local temperature defined
using a thermometer and thus verifies the same thermodynamical
properties of the latter.

This paper is organized as follows. In Sec. \ref{model}, we present
the model and summarize the theoretical treatment. In
Sec. \ref{temperatures} we review three definitions of temperature
addressed in recent works\cite{cal,cal2}. In Sec. \ref{correlations}
we derive general expressions for current-current correlation
functions and an explicit expression for the zero-frequency noise
within the Keldysh Green's functions formalism. In Sec. \ref{results} we
present numerical results for a particular system. Section
\ref{conclusions} is devoted to discussion and conclusions. We give
some details of the calculation in the Appendix.

\section{Model and theoretical treatment}\label{model}

\begin{figure}
\centering
\includegraphics[width=80mm,angle=0,clip]{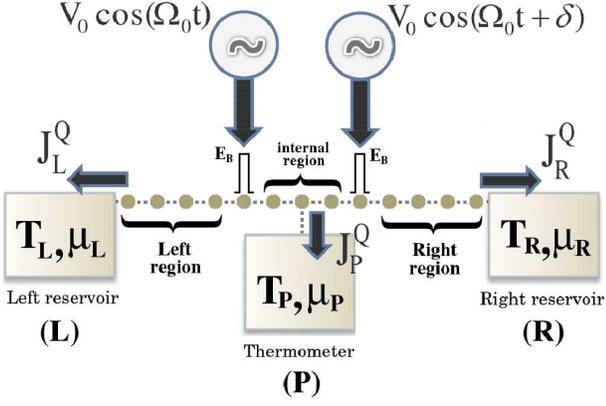} {\small {} }
\caption{{ Scheme of the setup. The central device is a wire with two
    barriers of height $E_B$ connected by its ends to two reservoirs
    ($L$ and $R$). The third reservoir ($P$) represents the probe,
    which consists of a macroscopic system weakly coupled to a given
    point of the central device. In this setup, transport is induced
    by two oscillating ac fields (both with the same amplitude and
    frequency but with a phase lag) applied at the points where the
    barriers are located. The Left and Right regions depicted in this 
    scheme are related to the heat current that flows into the
    respective reservoirs.\cite{cal2,cal3}}}
\label{setup}
\end{figure}

In Fig. \ref{setup} we display the same setup as in
Refs. \onlinecite{cal} and \onlinecite{cal2} representing a quantum
driven system, with the Hamiltonian $H_{sys}(t)$, connected to a
probe characterized by $H_P$. The total system is then described by
\be
H(t) = H_{sys}(t) + H_{cP} + H_P,
\ee
with $H_{cP}$ implementing the local coupling between the system and
the probe.  The Hamiltonian corresponding to the driven system can in
turn be written as
\be
H_{sys}(t) = H_L + H_{cL} + H_C(t) + H_{cR} + H_R,
\ee
where $H_C(t)$, $H_L$ and $H_R$ stand for the Hamiltonians of the
central part and left and right reservoirs, coupled among
themselves via the Hamiltonians $H_{cL}$ and $H_{cR}$.

The Hamiltonian describing the central system ($C$) contains the ac
fields and can be written as $H_C(t) = H_0 + H_V(t)$.  We assume that
$H_0$ is a Hamiltonian for non-interacting electrons while $H_V(t)$ is
harmonically time dependent with a fundamental driving frequency
$\Omega_0$. We leave further details of the model undetermined as much
of the coming discussion is model independent.

All three reservoirs (left, right and the probe) are modeled by
systems of non-interacting electrons with many degrees of freedom,
i.e., $H_\alpha = \sum_{k\alpha} \varepsilon_{k\alpha}
c^\dagger_{k\alpha} c_{k\alpha}$, where $\alpha=L,R,P$. The
corresponding contacts are described by $H_{c\alpha} = w_{c\alpha}
\sum_{k\alpha}(c^\dagger_{k\alpha} c_{l\alpha} + c^\dagger_{l\alpha}
c_{k\alpha})$, where $l\alpha$ denotes the coordinate of $C$ where the
reservoir $\alpha$ is connected. As in previous works,
\cite{cal,cal2,cal3,fourpoint,fourpointfed} we consider non-invasive
probe and we treat $w_{cP}$ at the lowest order of perturbation theory
when necessary.

We will analyze the out-of-equilibrium dynamics of this system within
the Schwinger-Keldysh Green's functions formalism.  Within this
formalism, instead of the usual time-ordering operator used in
equilibrium theory a contour-ordering operator which orders
time-labels according to their order on the Keldysh contour is
introduced. The single particle propagator reads
\be \label{green-Def}
i G_{j,j'}(t,t') = \langle T_\mathcal{C} [ c_j (t)
  c^{\dagger}_{j'}(t') ] \rangle.
\ee
The contour-ordered Green's function contains four different functions
depending on where the times $t$ and $t'$ are over the Keldysh
contour.\cite{keldysh} It is easy to see that they are not all
independent. We then consider the {\em lesser}, {\em greater} and {\em
  retarded} Green's functions,
\begin{eqnarray}
i G^{<}_{j,j'}(t,t') &=& - \langle c^{\dagger}_{j'}(t') c_j(t) \rangle, 
\nonumber \\
i G^{>}_{j,j'}(t,t') &=&  \langle c_j(t) c^{\dagger}_{j'}(t') \rangle, 
\nonumber \\
i G^R_{j,j'}(t,t')&=& \Theta(t-t') \langle \left[
  c_j(t),c^{\dagger}_{j'}(t') \right]_+  
\rangle ,
\label{green}
\end{eqnarray}
where $[,]_+$ denote the anticommutator of the fermionic operators,
$\langle ...\rangle$ is the quantum statistical average and the
indexes $j,j'$ denote spatial coordinates of the system. These Green's
functions can be evaluated after solving the Dyson equations.  

In this work we will focus on current-current correlation
functions. The current in reservoir $\alpha$ at time $t$ is defined by
the operator\cite{liliflo}
\be \label{current-op}
\hat{J}_\alpha(t) = i w_{c \alpha} \sum_{k \alpha} \left( \hat{c}_{k
  \alpha}^\dagger (t) \hat{c}_{l \alpha}(t) - \hat{c}_{l
  \alpha}^\dagger(t) \hat{c}_{k \alpha} (t) \right),
\ee
which obeys bosonic commutation rules. The ensuing connected
contour-ordered propagator reads in this case
\be \label{cc-Def}
 i C_{\alpha \beta} (t,t') = \langle T_\mathcal{C} [ \hat{J}_\alpha (t)
    \hat{J}_\beta (t') ] \rangle - \langle \hat{J}_\alpha(t) \rangle
  \langle \hat{J}_\beta(t') \rangle,
\ee
while the {\em lesser}, {\em greater} and {\em retarded} Green's
functions are
\ba \label{cc-Def2}
 i C^<_{\alpha \beta} (t,t') & = & \langle \hat{J}_\beta (t')
 \hat{J}_\alpha (t) \rangle - \langle \hat{J}_\alpha(t) \rangle
  \langle \hat{J}_\beta(t') \rangle,
\nonumber \\
 i C^>_{\alpha \beta} (t,t') & = & \langle \hat{J}_\alpha (t)
    \hat{J}_\beta (t') \rangle - \langle \hat{J}_\alpha(t) \rangle
  \langle \hat{J}_\beta(t') \rangle,
\nonumber \\
i C_{\alpha \beta}^R(t,t') & = & \Theta(t-t') \langle \left[
  \hat{J}_\alpha (t), \hat{J}_\beta (t') \right]_- \rangle.
\ea
where $[,]_-$ denote the commutator of the currents.

For the case of harmonic driving it is convenient to use the
Floquet-Fourier representation of the Green's functions:
\cite{liliflo}
\begin{equation}\label{floquet}
 A_{j,j'}(t, t-\tau) = \sum_{k=-\infty}^{\infty}
 \int_{-\infty}^{\infty} \frac{d\omega}{2 \pi} 
 e^{-i (k \Omega_0 t + \omega \tau)}  A_{j,j'}(k,\omega).
\end{equation}
where $A$ stands for single-particle (\ref{green-Def}) or
current-current (\ref{cc-Def}) propagators. 

In general the {\em Keldysh} and {\em retarded} Green's functions, can
be expressed in terms of the {\em lesser} and {\em greater} Green's
functions via
\ba
A_{j,j'}^K(t,t') & = & A_{j,j'}^>(t,t') +
A_{j,j'}^<(t,t'), 
\nonumber \\
A_{j,j'}^R(t,t') & = & \Theta(t-t') \left[
  A_{j,j'}^>(t,t') - A_{j,j'}^<(t,t')\right] .
\ea

From the definition given in Eq. (\ref{floquet}) it is straightforward
to see that the Floquet-Fourier components of these functions can be
written as  
\ba \label{greenFF}
A_{j,j'}^K(k,\omega) & = & A_{j,j'}^>(k,\omega) +
A_{j,j'}^<(k,\omega), 
\nonumber \\
A_{j,j'}^R(k,\omega) & = & i \int_{-\infty}^\infty \frac{d\omega'}{2\pi}
\frac{A_{j,j'}^>(k,\omega') -
  A_{j,j'}^<(k,\omega')}{\omega 
  - \omega' + i 0^+}.
\nonumber \\
\ea

\section{Defining the temperature}\label{temperatures}
\subsection{Local temperature determined by a probe}

In Ref. \onlinecite{cal} we defined the local temperature ($T_{lP}$)
of the site $lP$ of the system as the value of the temperature of the
probe such that the time-averaged heat exchange between the central
system and the probe vanishes.

It can be shown\cite{liliheatpump} that, given $H_C(t)$ without
many-body interactions, the dc component of the heat current flowing
from the central system to the thermometer can be expressed as ($\hbar
= k_B = e = 1$)
\ba
& &J_P^Q = \sum_{\alpha=L,R,P} \sum_{k=-\infty}^{\infty}
\int_{-\infty}^{\infty} \frac{d\omega}{2 \pi} 
 \Big\{  [f_\alpha(\omega)-f_P(\omega_k)] \nonumber \\
& &   \times (\omega_k - \mu) \Gamma_P(\omega_k) \Gamma_\alpha(\omega)
 \left| G^R_{lP,l\alpha}(k,\omega)\right|^2 \Big\}, \label{jq}
\ea
where $\omega_k=\omega+k\Omega_0$, while $\Gamma_{\alpha}(\omega) = -2
\pi |w_{\alpha}|^2 \sum_{k \alpha} \delta(\omega-\varepsilon_{k
  \alpha})$ are the spectral functions characterizing 
the reservoirs ($\alpha = L, R, P$), and $f_\alpha(\omega)=
1/[e^{\beta_{\alpha}(\omega -\mu_{\alpha})}+1]$ is the Fermi
function, which depends on $T_{\alpha}=1/\beta_{\alpha}$ and
$\mu_\alpha $ respectively the temperature and the chemical potential
of the reservoir $\alpha$. 
Thus, the local temperature $T_{lP}$ corresponds to the solution of
the equation
\be \label{Tlocal}
J_P^Q(T_{lP}) = 0. 
\ee

In general, Eq. (\ref{Tlocal}) must be solved numerically, but under
certain conditions, an analytical expression can be found. In
particular, for the low temperature weak-driving adiabatic regime,
which corresponds to small amplitudes and frequencies of the driving
potential, and for $\Omega_0 \ll T$,\cite{cal}
\be \label{Tlocal-final}
T_{lP} = T \left[ 1 + \lambda^{(1)}_{lP}(\mu) \Omega_0 \right], 
\ee
where
\be \label{lambda-n}
\lambda_{l}^{(n)}(\omega)=  \frac{1}{ \sum_{k=-1}^{1}
  \varphi_{l}(k,\omega)} 
\sum_{k=-1}^{1} (k)^{n+2} \frac{d^n[ \varphi_{l}(k,\omega)]}{d
  \omega^n} , 
\ee
\be \label{varphi}
\varphi_{l}(k,\omega)  =   \sum_{\alpha= L,R}
 \left|G^R_{l, l \alpha}(k,\omega)\right|^2 \Gamma_{\alpha}(\omega).
\ee 

An alternative definition of local temperature was discussed in
Ref. \onlinecite{cal2}, where the fact that the heat current is
related to the charge current was taken into account. Then, the local
temperature ($T^*_{lP}$) and the local chemical potential
($\mu^*_{lP}$) were defined from the condition of simultaneously
vanishing of the time-averaged charge and heat currents between the
probe and the system. That is
\be
\left\{ \begin{array}{rcl}
J^Q_P (T_{lP}^*,\mu_{lP}^*) & = & 0, \\\nonumber
J^e_P (T_{lP}^*,\mu_{lP}^*) & = & 0,
\end{array} \right.
\label{Tlocal2}
\ee
where (see Refs. \onlinecite{liliflo,fourpointfed})
\ba
& &J_P^e = \sum_{\alpha=L,R,P} \sum_{k=-\infty}^{\infty}
\int_{-\infty}^{\infty} \frac{d\omega}{2 \pi} 
 \Big\{  [f_\alpha(\omega)-f_P(\omega_k)] \nonumber \\
& &   \times 
\Gamma_P(\omega_k) \Gamma_\alpha(\omega) \left|
G^R_{lP,l\alpha}(k,\omega)\right|^2 \Big\}, \label{je} 
\ea
is the dc component of the charge current flowing through the contact
between the system and the probe.

The simultaneous equations given in Eq. (\ref{Tlocal2}) can be solved
numerically for any situation, but an analytical expression can be
found within the low temperature weak-driving adiabatic
regime, when $\Omega_0 \ll T$, and leads to $T^*_{lP} = T_{lP}$,
given in Eq. (\ref{Tlocal-final}).

\subsection{Effective temperature from a single-particle
  fluctuation-dissipation relation (FDR)}

For systems in equilibrium, the fluctuation-dissipation theorem
establishes a relation between the Keldysh (correlation) and the
retarded Green's functions. In Ref. \onlinecite{cal}, we defined a
local FDR involving single-particle Green's functions from which an
effective temperature for the site $l$ ($T_{l}^{eff} =
1/\beta_{l}^{eff}$) can be extracted,
\ba \label{Teff}
iG^K_{l,l}(0,\omega)-iG^K_{l,l}(0,\mu) & = &
\tanh \left[ \frac{\beta^{eff}_{l} (\omega-\mu)}{2} \right] 
\overline{\varphi}_{l}(\omega), \label{fdr} 
\nonumber \\ 
\ea 
with $\overline{\varphi}_{l}( \omega)=-2 \, \mbox{Im}
[G_{l,l}^R(0,\omega)]= \sum_k \varphi_l(k,\omega_{-k})$.  In general,
Eq. (\ref{Teff}) defines an effective temperature that might depend on
$\omega$, so the limit $\omega \rightarrow \mu$ is taken. An extra
term is added to the lhs of Eq. (\ref{Teff}) because the rhs is always
zero at $\omega = \mu$ but $G^K_{l,l}(0,\mu)$ is not necessarily
zero in an arbitrary out-of-equilibrium situation.

Within the low temperature weak-driving adiabatic regime, when
$\Omega_0 \ll T$, we showed\cite{cal} that $T^{eff}_{lP} =
T_{lP}$. Then, the conclusion of our previous investigations is that
for the weak driving adiabatic regime the effective temperature
defined from a single-particle FDR coincides with that determined by a
thermometer.
 
\section{Current-current correlation functions and Effective
  Temperature} 
\label{correlations}

\subsection{A non-equilibrium fluctuation-dissipation
  relation} \label{Teff2-sec}

We analyze the role of effective temperatures ($T^{eff*}$) from a FDR
in the framework of two-particle correlation functions. As {\em a
  priori} they are not necessarily the same as the effective
temperatures defined above we use an asterisk to refer to them.  We
will focus on current-current correlation functions since they are
more easily accessible from an experimental point of view. We are
particularly interested in a local relation, that is both currents
evaluated at the same point.

As in the case of the single-particle FDR we focus on the dc
components of the correlation functions to define the effective
temperature. This corresponds to assuming that an equilibrium-like FDR
holds for the $k=0$ Floquet component with $\beta^{eff*}$ playing the
role of the inverse of temperature,
\be \label{fdr2}
C_{\alpha \alpha}^R(0,\omega) = i \int_{-\infty}^\infty
\frac{d\omega'}{2\pi} \frac{C_{\alpha \alpha}^K(0,\omega')}{\omega -
  \omega' + i 0^+} \tanh
\left[\frac{\beta^{eff*}_{l\alpha}\omega'}{2}\right].  
\ee

An equivalent expression for the FDR given in Eq. (\ref{fdr2}) is
obtained by considering the imaginary part, which leads to
\be \label{Teff2}
i C_{\alpha \alpha}^K(0,\omega) = 
 \coth \left[ \frac{\beta^{eff*}_{l\alpha} \omega}{2} \right]
\overline{\varphi}^*_\alpha(\omega), 
\ee
where $\overline{\varphi}^*_\alpha(\omega) = - 2 \, \mbox{Im} \left[
  C_{\alpha \alpha}^R(0,\omega) \right]$. (Notice that the real part is
simply derived by means of Kramers-Kronig relations.) As in the case
of the single-particle FDR in Eq. (\ref{Teff}), Eq. (\ref{Teff2})
defines an effective temperature that might depend on $\omega$, so the
limit $\omega \rightarrow 0$ is taken.

It is important to notice the similarity of this expression with the
one shown in Eq. (\ref{Teff}) for single-particle Green's functions
(fermionic operators). In this case the hyperbolic tangent is replaced
by an hyperbolic cotangent due to the bosonic statistic of current
operators. 

\subsection{Current-current correlation and noise}

Although we are more interested in the case of local current
correlations, let us start by considering the more general case of
correlation at different points.  If we consider two reservoirs
($\alpha$ and $\beta$) and two times (an absolute time $t$ and a
relative time $\tau$) we can define the correlation function of
currents as
\be \label{corr-cc}
P_{\alpha \beta}(t,t-\tau) = \frac{1}{2} \langle \Delta \hat{J}_\alpha(t)
\Delta \hat{J}_\beta(t-\tau) + \Delta \hat{J}_\beta(t-\tau) \Delta
\hat{J}_\alpha(t) \rangle,
\ee
where $\Delta \hat{J}_\alpha(t) = \hat{J}_\alpha(t) - \langle
\hat{J}_\alpha(t) \rangle$.

With the definition of the contour-ordered current-current correlation
function given in Eq. (\ref{cc-Def}), the correlation function of
currents given in Eq. (\ref{corr-cc}) can be expressed as 
\ba \label{corr-cc2}
P_{\alpha \beta} (t,t-\tau) & = & \frac{i}{2} \left( C_{\alpha
  \beta}^>(t,t-\tau) 
+ C_{\alpha \beta}^< (t,t-\tau) \right) \nn \\
& = & \frac{i}{2} C_{\alpha \beta}^{K} (t,t-\tau). 
\ea
If instead of a symmetrized current-current correlation we are
interested in a nonsymmetrized one,
\be
P^{ns}_{\alpha \beta}(t,t-\tau) = \langle \Delta \hat{J}_\alpha(t)
\Delta \hat{J}_\beta(t-\tau) \rangle,
\ee
the correlation becomes
\be
P^{ns}_{\alpha \beta} (t,t-\tau) = i C_{\beta \alpha}^< (t-\tau,t). 
\ee
In this work we will give results for the symmetrized correlation only
but it is straightforward to obtain the results for the nonsymmetrized
one. 

Since experimentally the noise spectrum is averaged over the absolute
time $t$, the relevant quantity here is
\be
\mathcal{P}_{\alpha \beta}(\omega) = 2 \int d\tau \langle P_{\alpha
  \beta}(t,t-\tau) \rangle_t e^{i \omega \tau}
\ee
where $\langle \ldots \rangle_t$ denotes the time average. From the
definition of the Floquet-Fourier components given in
Eq. (\ref{floquet}) it is easy to see that
\be
\mathcal{P}_{\alpha \beta}(\omega) = i C_{\alpha \beta}^K(0,\omega).
\ee
Hence, the only relevant Floquet-Fourier component is the one with
$k=0$.

As we are considering non-interacting electrons the
contour-ordered propagator given in Eq. (\ref{cc-Def}) can be
exactly evaluated in terms of single-particle propagators
(\ref{green-Def}). Using Wick's theorem (see the Appendix for the
details), this contour-ordered function can be written as
\ba \label{C-calc1}
  i C_{\alpha \beta} (t,t') & = & - w_{c \alpha} w_{c \beta} \sum_{k
    \alpha, k \beta} 
  \left\{ G_{l\beta,k\alpha}(t',t) G_{l\alpha,k\beta}(t,t') \right.
    \nonumber \\ & &
    - G_{k\beta,k\alpha}(t',t) G_{l\alpha,l\beta}(t,t')
    \nonumber \\ & &
    - G_{l\beta,l\alpha}(t',t) G_{k\alpha,k\beta}(t,t')
    \nonumber \\ & &
    \left. + G_{k\beta,l\alpha}(t',t) G_{k\alpha,l\beta}(t,t') \right\} .
\ea

In the Appendix we show the detailed calculation leading from this
expression to the Floquet-Fourier components $C^\gtrless_{\alpha
  \beta}(0,\omega)$. Here we only reproduce the results for two cases
of particular interest.

The first case is the zero-frequency limit of $C^K_{\alpha
  \beta}(0,\omega)$, which reads
\be
P_{\alpha \beta} \equiv \frac{i}{2} C^K_{\alpha \beta}(k=0,\omega=0) =
\delta_{\alpha \beta} P_\alpha + P_{\alpha \ne \beta},
\ee
where
\ba
P_\alpha & = & \int_{-\infty}^\infty \frac{d\omega'}{2\pi}
\Gamma_\alpha(\omega') \sum_{k=-\infty}^\infty \sum_\gamma
\Gamma_\gamma(\omega'_k) f_{\alpha \gamma}(\omega',\omega'_k) 
\nonumber \\ && 
\times |G^R_{l\alpha,l\gamma}(-k,\omega'_k)|^2,
\nonumber \\
P_{\alpha \ne \beta} & = & - \frac{1}{2} \int_{-\infty}^\infty
\frac{d\omega'}{2\pi} 
\Gamma_{\alpha}(\omega') \sum_{k=-\infty}^\infty
\Gamma_\beta(\omega'_k) \Big\{ 
f_{\alpha \beta}(\omega',\omega'_k)
\nonumber \\ && 
\times \mbox{Re} \Big[ G^R_{l\beta,l\alpha}(k,\omega')
  G^R_{l\alpha,l\beta}(-k,\omega'_k) \Big] 
\nonumber \\ && 
- 2 \sum_{k=-\infty}^\infty \sum_\gamma \Gamma_\gamma(\omega'_{k'})
f_{\alpha \gamma} (\omega',\omega'_{k'}) \mbox{Im} \Big[
  G^R_{l\beta,l\alpha}(k,\omega')
\nonumber \\ && 
\times G^R_{l\alpha,l\gamma}(-k',\omega'_{k'})
G^R_{l\beta,l\gamma}(k-k',\omega'_{k'})^* \Big]
\nonumber \\ && 
+ G^>_{l\beta,l\alpha}(k,\omega')G^<_{l\beta,l\alpha}(k,\omega')^* \Big\}
\nonumber \\ && 
+ \Big\{ \mbox{same with } \alpha \leftrightarrow \beta \Big\},
\ea
being
\be
f_{\alpha \beta}(\omega,\omega') = f_\alpha(\omega) 
(1 - f_\beta(\omega')) + f_\beta(\omega') (1 - f_\alpha(\omega)).
\ee

It is important to notice that this is a general result for
multiterminal quantum driven systems and the sum over $\gamma$ extends
over all reservoirs connected to the central system. For this work we
chose a two terminal system, but this result is completely general as
no assumption concerning the reservoirs was made in the calculation.

At this point it is interesting to compare with previous results
obtained within the scattering matrix formalism (see
Ref. \onlinecite{ButtikerNoise}). In order to do so, we need to assume
that all reservoirs are at equal temperature and chemical potential
(unbiased pump). We split the zero frequency noise into two
contributions 
\be
P_{\alpha \beta} \equiv P_{\alpha \beta}^{(th)} + P_ {\alpha \beta}^{(sh)}, 
\ee
where
\ba \label{zero-freq-noise}
P_{\alpha \beta}^{(th)} & = & \int_{-\infty}^\infty \frac{d\omega'}{2\pi}
f(\omega') (1 - f(\omega')) \Gamma(\omega') \sum_{k=-\infty}^\infty
\Gamma(\omega'_k) 
\nonumber \\ &&
\times \bigg\{ \delta_{\alpha \beta} \sum_\gamma \Big(
|G^R_{l\alpha,l\gamma}(k,\omega')|^2 +
|G^R_{l\alpha,l\gamma}(-k,\omega'_k)|^2  
\Big) 
\nonumber \\ &&
- |G^R_{l\alpha,l\beta}(k,\omega')|^2 -
|G^R_{l\beta,l\alpha}(k,\omega')|^2 \bigg\}, 
\nonumber \\
P^{(sh)}_{\alpha \beta} & = & \int_{-\infty}^\infty \frac{d\omega'}{2\pi}
\Gamma(\omega') \sum_{k=-\infty}^\infty \Gamma(\omega'_k) \bigg\{
\delta_{\alpha \beta} 
\left( f(\omega') - f(\omega'_k) \right)^2
\nonumber \\ &&
\times |G^R_{l\alpha,l\gamma}(k,\omega')|^2 - f(\omega')^2 \Big(
|G^R_{l\alpha,l\beta}(k,\omega')|^2 
\nonumber \\ &&
+ |G^R_{l\beta,l\alpha}(k,\omega')|^2 \Big) + 2 f(\omega')f(\omega'_k)
\mbox{Re} \Big[ G^R_{l\beta,l\alpha}(k,\omega')
\nonumber \\ &&
\times G^R_{l\alpha,l\beta}(-k,\omega'_k) \Big] - 2 \mbox{Re} \Big[ \Big(
f(\omega') G^R_{l\beta,l\alpha}(k,\omega')
\nonumber \\ &&
 - f(\omega'_k) G^R_{l\alpha,l\beta}(-k,\omega'_k)^* \Big)
 G^<_{l\beta,l\alpha}(k,\omega')^* \Big]
\nonumber \\ &&
- |G^<_{l\beta,l\alpha}(k,\omega')|^2 \bigg\}.
\ea

The first term $P^{(th)}_{\alpha \beta}$ is the Nyquist-Johnson noise
while $P^{(sh)}_{\alpha \beta}$ is the shot noise. Using the relation
between the Floquet S-matrix and Green's functions \cite{liliflo}
\ba
S_{F,\alpha \beta}(\omega_m,\omega_n) & = & \delta_{\alpha \beta}
\delta_{n,m} - i \sqrt{\Gamma(\omega_m)\Gamma(\omega_n)}
\nonumber \\ &&
\times G^R_{l\alpha,l\beta}(m-n,\omega_n),
\ea
it is easy to show that the result given in
Eq. (\ref{zero-freq-noise}) coincides with the one obtained using the
Floquet S-matrix formalism in Ref. \onlinecite{ButtikerNoise}.

The other case of interest is the one in which $\alpha=\beta=P$,
i.e. we concentrate in current fluctuations of the probe. Using the
fact that the probe is noninvasive, we only keep terms to the lowest
order in the coupling $w_{cP}$ between the system and the
thermometer, 

\ba \label{ccK}
i C_{PP}^K(0,\omega) & = & \Gamma_P \int_{-\infty}^\infty
\frac{d\omega'}{2\pi} \sum_{k=-\infty}^\infty\sum_{\gamma=L,R}
\Gamma_\gamma(\omega') \Big\{ f_\gamma(\omega')
\nonumber \\
& & \times \big[2 - f_P(\omega'_{k} + \omega) - f_P(\omega'_{k} -
  \omega) \big] 
\nonumber \\
&& + \big[f_P(\omega'_{k} + \omega) + f_P(\omega'_{k} - \omega)\big]
\nonumber \\
&& \times (1 - f_\gamma(\omega')) \Big\} |
G_{lP,l\gamma}^R(k,\omega') |^2. 
\ea

On the other hand we need $\overline{\varphi}^*_P(\omega)$, which
is    
\ba \label{ccR}
\overline{\varphi}^*_P(\omega)
& = & \Gamma_P \int_{-\infty}^\infty \frac{d\omega'}{2\pi}
\sum_{k=-\infty}^\infty \sum_{\gamma=L,R} 
\Gamma_\gamma(\omega') 
\nonumber \\
& & \times \big[f_P(\omega'_{k} - \omega) - f_P(\omega'_{k} +
  \omega) \big]
\nonumber \\
& & \times | G_{lP,l\gamma}^R(k,\omega') |^2.
\ea

The functions entering Eqs. (\ref{ccK}) and (\ref{ccR}) are the ones
involved in the definition of effective temperature given in
Eq. (\ref{Teff2}).

\section{Results}\label{results}

In this section we present results for a central device consisting of
non-interacting electrons in a one-dimensional lattice:
\be
H_0= -w \sum_{l,l^{\prime}} (c^\dagger_{l} c_{l^{\prime}} + H.c.),
\ee
where $w$ denotes a hopping matrix element between neighboring
positions $l,l^{\prime}$ on the lattice. The driving term is chosen as
\be \label{hv}
 H_V(t)= \sum_{j=1}^2 e V_j(t) c^\dagger_{lj} c_{lj} ,
\ee
with $V_j(t)= E_B + V_0 \cos( \Omega_0 t + \delta_j)$, $lj$ being the
positions where two oscillating fields with frequency $\Omega_0$ and
phase-lag $\delta$ are applied. 
This defines a simple model for a quantum pump where two ac gate
voltages are applied at the walls of a quantum
dot. \cite{liliflo,adia,pump}   

\subsection{Equivalence between effective and local temperature at
  weak driving}

As in Refs. \onlinecite{cal} and \onlinecite{cal2} we are interested
in the weak driving regime, which corresponds to a situation where the
ac voltage amplitudes are lower than the kinetic energy of the
electrons in the structure and the driving frequency is much smaller
than the inverse of the dwell time of these electrons. We have shown
that in this regime the local temperature defined from
Eq. (\ref{Tlocal}), with the chemical potential of the probe fixed, is
identical to the local temperature defined from Eq. (\ref{Tlocal2}),
where the chemical potential of the probe has to be determined in
order to satisfy both equations, and it is also identical to an
effective temperature defined from a local fluctuation-dissipation
relation of single-particle Green's functions (see Eq. (\ref{Teff})).

We now turn our attention to the effective temperature $T^{eff*}$
defined in Eq. (\ref{Teff2}), involving current-current correlation
functions. The correlation functions given in Eqs. (\ref{ccK}) and
(\ref{ccR}) depend on the temperature $T_P$ and the chemical potential
$\mu_P$ of the probe via the Fermi function $f_P$. Thus, the effective
temperature $T^{eff*}$, so calculated, also depends on $T_P$ and
$\mu_P$. There are many possible reasonable choices for the latter
quantities. In this subsection we will concentrate in only one choice
and leave for the next subsection the analysis of other
possibilities. We choose $\mu_P$ equal to the chemical potential $\mu$
of the reservoirs and $T_P$ equal to the local temperature $T_{lP}$,
i.e. the one for which the heat flow between the system and the probe
vanishes.

\begin{figure}
\centering
\includegraphics[width=80mm,angle=0,clip]{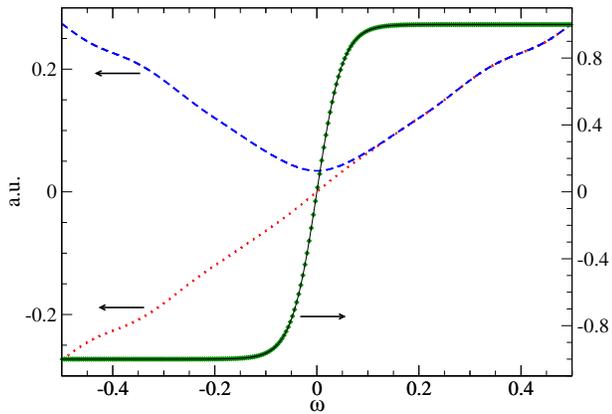} {\small {} }
\caption{{ (Color online) Current-current correlation functions
    $\overline{\varphi}_P^*(\omega)$ (dotted red),
    $iC^K_{PP}(0,\omega)$ (dashed blue), their quotient (green
    diamonds), and $\tanh [\beta^{eff*} \omega/2]$ (solid black) as a
    function of $\omega$. The reservoirs have chemical potential
    $\mu=0.2$ and temperature $T=0.025$. The driving frequency is
    $\Omega_0=0.01$, the amplitude is $V_0=0.05$ and $E_B = 0.2$.}}
\label{tanh}
\end{figure}

In Fig. \ref{tanh} we show a typical plot for $iC_{PP}^K(0,\omega)$,
$\overline{\varphi}_P^*(\omega)$ and their ratio as a function of
$\omega$. According to the definition of effective temperature given
in Eq. (\ref{Teff2}), the derivative of this ratio at $\omega = 0$
corresponds to $\beta^{eff*}/2$. This derivative is calculated
numerically. In the same figure we plot $\tanh\left[
  \beta^{eff*} \omega/2 \right]$ and we see that the quotient
$\overline{\varphi}_P^*(\omega)/iC_{PP}^K(0,\omega)$ is well fitted by
a FDR-type relation for a reasonably large frequency interval.

\begin{figure}
\centering
\includegraphics[width=80mm,angle=0,clip]{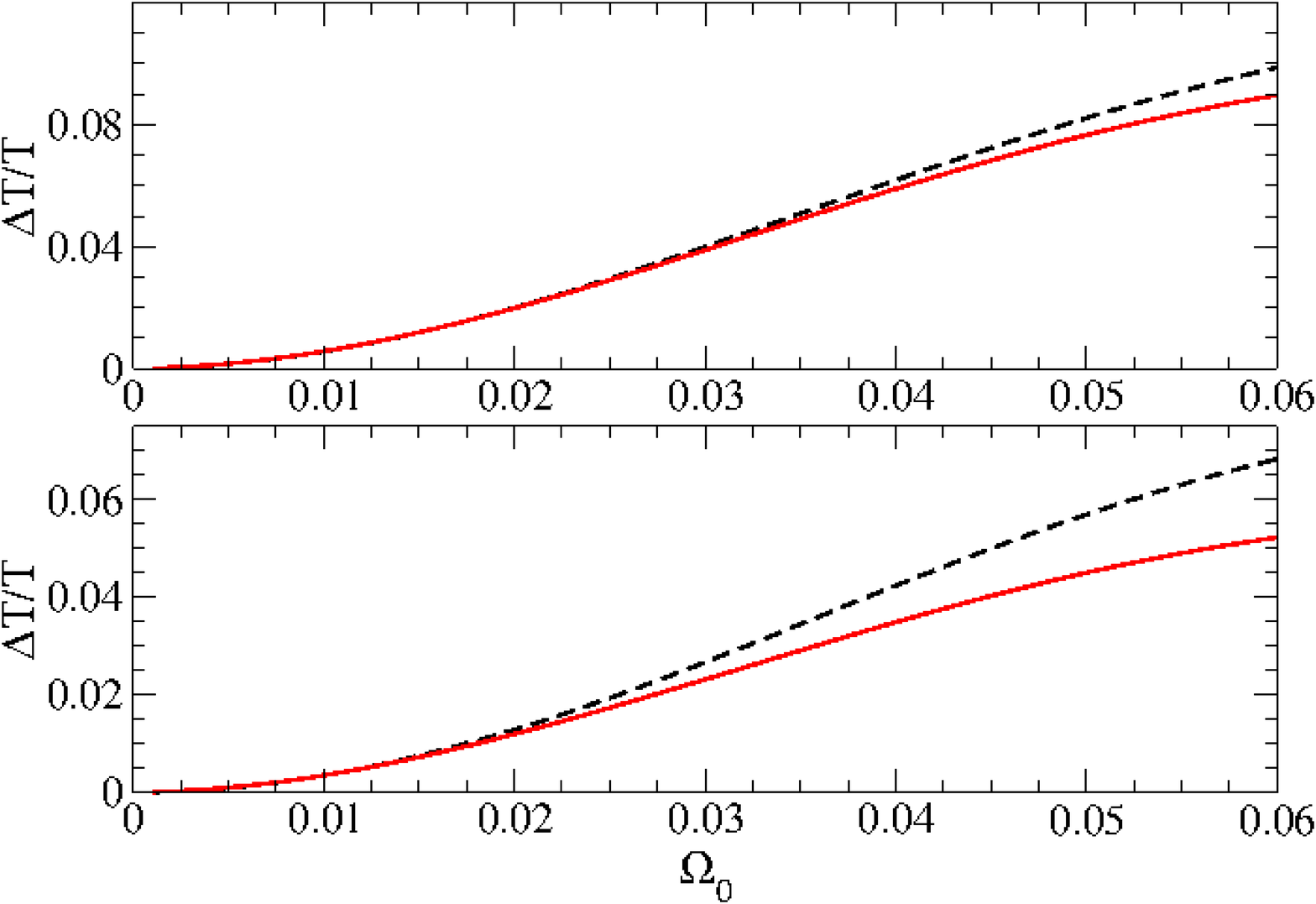} {\small {} }
\caption{{ (Color online) Local temperature $T_{lP}$ (dashed black)
    and effective temperature $T^{eff*}_{lP}$ (solid red) for the site
    $lP = lL$ (i.e. the site connected to the left reservoir) as a
    function of driving frequency $\Omega_0$. The reservoirs have
    chemical potential $\mu=0.2$. The upper panel corresponds to
    $T=0.016$, while the lower panel corresponds to $T=0.005$.}}  
\label{Omega0}
\end{figure}

In Fig. \ref{Omega0} we show the behavior of the effective temperature
$T^{eff*}$ and the local temperature $T_{lP}$, calculated for the site
connected to the left reservoir, as a function of the driving
frequency $\Omega_0$ for two different temperatures of the
reservoirs. This analysis can be done for any site of the central
system but we chose this particular site because its local temperature
determine the heat current that flows into the left
reservoir.\cite{cal2} Results for any other site of the central system
are similar. In Fig. \ref{Omega0} the upper panel corresponds to
$T=0.016$, while the lower corresponds to $T=0.005$. We see that both
ways of defining the temperature coincide at low frequencies. This
supports the idea that, for a given temperature $T$ of the reservoirs,
$T^{eff*}$ is a {\em bona fide} temperature within the low driving
regime. As we can see from Fig. \ref{Omega0}, the higher the
temperature $T$ of the reservoirs, the broader the region of low
driving frequency $\Omega_0$ in which the two definitions of the
temperature agree.

\begin{figure}
\centering
\includegraphics[width=80mm,angle=0,clip]{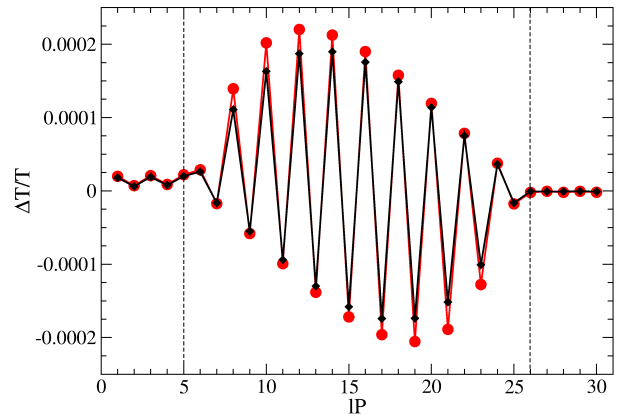} {\small {} }
\caption{{
(Color online) Local temperature $T_{lP}$ (black diamonds) and
    effective temperature $T_{lP}^{eff*}$ (red circles) along a
    one-dimensional model of $N=30$ sites with two ac fields operating
    with a phase lag of $\delta=\pi/2$ at the positions indicated by
    dotted lines. The system is in contact with reservoirs with
    chemical potentials $\mu=0.2$ and temperature $T=0.02$. The
    driving frequency is $\Omega_0=0.001$, the amplitude is $V_0
    =0.05$ and $E_B = 0.2$. }} 
\label{sitios}
\end{figure}

As we mentioned earlier, the definition given in Eq. (\ref{Teff2}) can
be used to calculate the effective temperature in each site of the
central system. In Fig. \ref{sitios} we show the comparison between
the local temperature $T_{lP}$ and the effective temperature
$T^{eff*}$ all along the sample. The values of $T_{lP}$ and $T^{eff*}$
are plotted for each point of the linear chain, for $T_L = T_R = T =
0.02$, $\mu_L = \mu_R = \mu = 0.2$ and a particular low value of
$\Omega_0 = 0.001$. We can see that there is a good agreement between
the two temperatures along the whole structure and an almost perfect
agreement within the ``Left'' and ``Right'' regions (defined in
Fig. \ref{setup}), which are the ones from where we can determine the
heat flow between the system and each one of the reservoirs (see
Refs. \onlinecite{cal2},\onlinecite{cal3}). It is also important to
notice the existence of $2k_F$ Friedel-like oscillations, $k_F$ being
the Fermi vector of the electrons leaving the reservoirs. These
oscillations are an indication of quantum interference. They were
previously reported for exactly the same setup we study in this work
\cite{cal,cal2} and also predicted in other mesoscopic systems under a
stationary driving. \cite{dubi-diventra}

\subsection{Different choices of $T_P$ and $\mu_P$}

The effective temperature $T^{eff*}$ depends on the values of $T_P$
and $\mu_P$ (respectively the temperature and chemical potential of
the probe). The choice analyzed in the previous section was $\mu_P$
equal to the chemical potential $\mu$ of the reservoirs and $T_P$
equal to the local temperature $T_{lP}$. We will call this choice Case
I. Another suitable choice (Case II) could be to choose $\mu_P = \mu$,
as in the previous case but $T_P$ such that $T^{eff*} = T_P$. A third
choice (Case III) could be to choose $T_P$ such that $T^{eff*} = T_P$
but at the same time $\mu_P = \mu_{lP}$ (the local voltage) in order
to have a vanishing charge current between the system and the probe at
that temperature. In this work we will only deal with these three
possibilities.

\begin{figure}
\centering
\includegraphics[width=80mm,angle=0,clip]{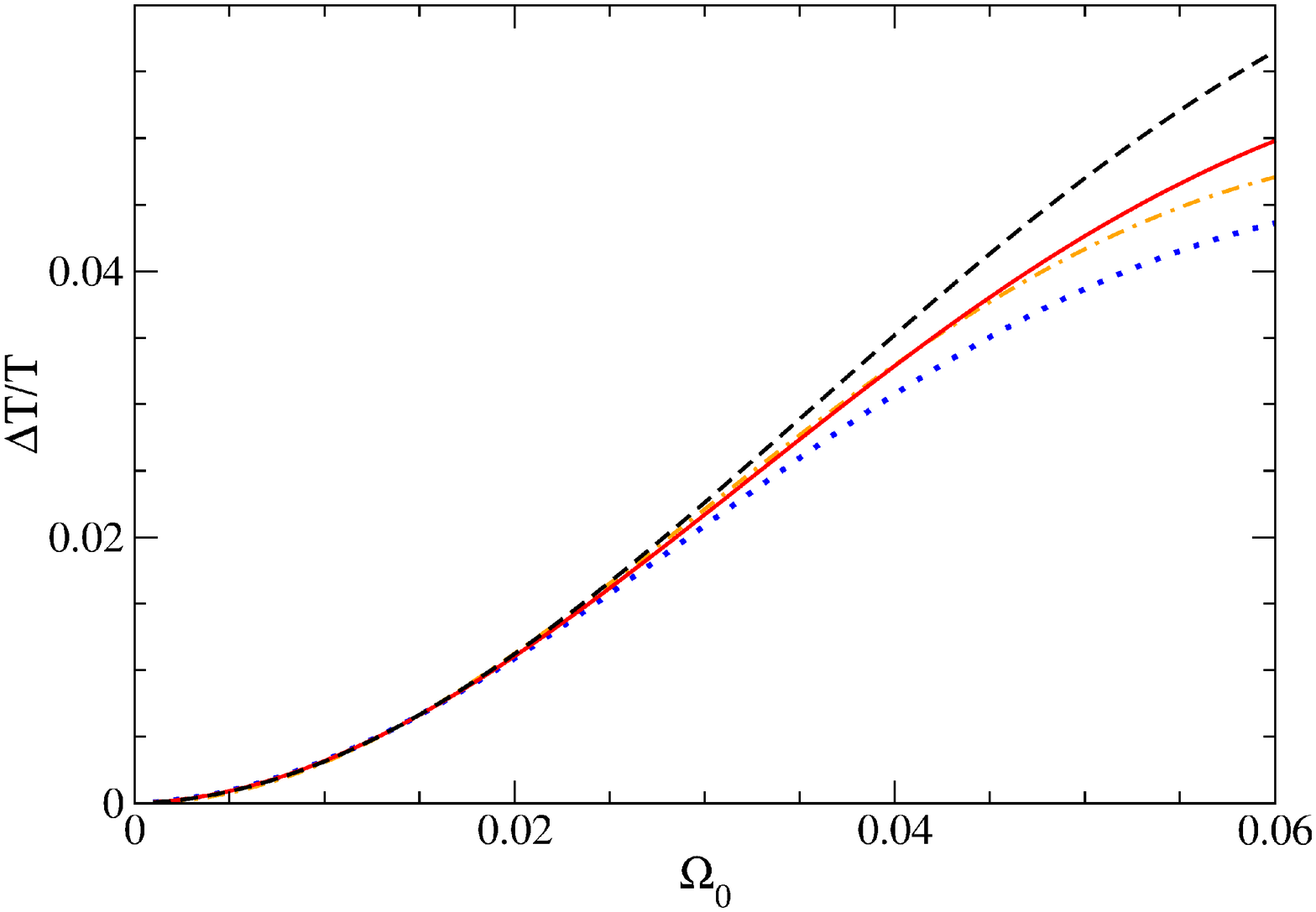} {\small {} }
\caption{{ (Color online) Local temperature $T_{lP}$ (dashed black)
    and effective temperature $T^{eff*}_{lP}$ for Case I (solid red),
    Case II (dotted blue) and Case III (dashed and dotted orange) for
    the site $lP = lL$ as a function of driving frequency
    $\Omega_0$. The reservoirs have chemical potential $\mu=0.2$ and
    temperature $T=0.01$.}}
\label{comp}
\end{figure}

If Fig. \ref{comp} we show the three different effective temperatures
corresponding to the above mentioned cases together with the local
temperature as a function of the driving frequency $\Omega_0$ for a
given temperature of the reservoirs. As we can see, all three cases
give a good estimate of the local temperature in the regime of
interest (i.e. low driving frequencies). This behavior supports the
robustness of the definition of the local temperature from a FDR.

\section{Summary and Conclusions}\label{conclusions}
In this work we have calculated the current-current correlation
functions for quantum driven systems and found an explicit expression
for the zero-frequency noise within the Schwinger-Keldysh Green's
functions formalism. In the particular case of multiterminal unbiased
quantum driven systems our result is in agreement with previous
results obtained within the scattering matrix
approach.\cite{ButtikerNoise} For non-interacting systems both
descriptions agree, while the Green's functions has the advantage of
providing a systematic framework for the study of interacting systems.

We have also defined an effective temperature from a local
fluctuation-dissipation relation for current-current correlation
functions and showed that for low frequencies it coincides with the
local temperature defined with a thermometer and from a FDR at the
level of single-particle propagators.  This result opens the
possibility of using current-current correlation in real experiments,
in order to define the local temperature of a driven sample.

%%%%%%%%%%%%%%%%%%%%%%%%%%%%%%%%%%%%%%%%%%%%%%%%%%%%%%%%%%%%%%%%%%%%%%%%%%%%%%%%%%%%%%%%%%%%%%%%%%%%%
%%%%%%%%%%%%%%%%%%%%%%%%%%%%%%%%%%%%%%%%%%%%%%%%%%%%%%%%%%%%%%%%%%%%%%%%%%%%%%%%%%%%%%%%%%%%%%%%%%%%

\begin{acknowledgments}
We thank M. B\"uttiker and L. Cugliandolo for valuable discussions. We
acknowledge support from CONICET, ANCyT, UBACYT, Argentina and
J. S. Guggenheim Memorial Foundation (LA).
\end{acknowledgments}

\appendix*

\section{Analytical expressions for current-current correlation
  functions}

Using Eq. (\ref{current-op}), the connected contour-ordered
current-current correlation (see Eq. (\ref{cc-Def})) can be written in
terms of electron operators as
\ba \label{C-calc0}
  i C_{\alpha \beta} (t,t') & = & - w_{c\alpha} w_{c\beta} \sum_{k\alpha,
    k\beta} \Big( D_{k\alpha,l\alpha,k\beta,l\beta} (t,t')
    \nonumber \\ & &
    - D_{k\alpha,l\alpha,l\beta,k\beta} (t,t')
    \nonumber \\ & &
    - D_{l\alpha,k\alpha,k\beta,l\beta} (t,t')
    \nonumber \\ & &
    + D_{l\alpha,k\alpha,l\beta,k\beta} (t,t') \Big),
\ea
where
\ba \label{D-Def}
D_{i,j,k,l} (t,t') & = & \langle T_\mathcal{C} [ \hat{c}_i^\dagger (t)
  \hat{c}_{j}(t) \hat{c}_k^\dagger (t') \hat{c}_l(t') ] \rangle 
  \nonumber \\ & &
  - \langle \hat{c}_i^\dagger (t) \hat{c}_{j}(t) \rangle 
  \langle \hat{c}_k^\dagger (t') \hat{c}_l(t') \rangle. 
\ea

Using Wick's theorem and the definition of the contour-ordered Green's
function (Eq. (\ref{green-Def})) we can rewrite Eq. (\ref{D-Def}) as
\be \label{D-calc}
D_{i,j,k,l} (t,t') = G_{l,i} (t',t) G_{j,k}(t,t').
\ee
The substitution of Eq. (\ref{D-calc}) into Eq. (\ref{C-calc0}) gives
the result shown in Eq. (\ref{C-calc1}).

The next step is to calculate the lesser and greater Green's
functions. In order to have the lesser (greater) Green's function we
need $t \in C_1$ and $t' \in C_2$ ($t \in C_2$ and $t' \in C_1$).  If
$t \in C_1$ and $t' \in C_2$, then
\be \label{D-lesser}
D^<_{i,j,k,l} (t,t') = G^>_{l,i} (t',t) G^<_{j,k}(t,t').
\ee

Using Eq. (\ref{D-lesser}) we can write the lesser Green's function as
\ba \label{C-lesser}
i C^<_{\alpha \beta} (t,t') & = & w_{c \alpha} w_{c \beta} \sum_{k
  \alpha, k \beta} 
\left\{ \left( G^>_{k\alpha,l\beta}(t,t')^* G^<_{l\alpha,k\beta}(t,t') 
\right) \right.
\nonumber \\ & &
- \left(
G^>_{k\alpha,k\beta}(t,t')^* G^<_{l\alpha,l\beta}(t,t')
\right)
\nonumber \\ & &
- \left(
G^>_{l\alpha,l\beta}(t,t')^* G^<_{k\alpha,k\beta}(t,t') 
\right)
\nonumber \\ & &
\left.
+ \left( G^>_{l\alpha,k\beta}(t,t')^* G^<_{k\alpha,l\beta}(t,t')
\right)
\right\}.
\ea
where the property $G^{<,>}_{j,j'}(t,t') = - G^{<,>}_{j',j}(t',t)^*$
was used. Notice that the last term is equal to the first
interchanging $<$ with $>$ and conjugating. The same situation arises
with the second and third terms. The result for the greater Green's
function is obtained by switching $<$ with $>$.

We are interested in the Floquet-Fourier components of this Green's
functions. So, according to Eq. (\ref{floquet}) we need to calculate
\be \label{C-lesser-floquet}
C^<_{\alpha \beta} (t,t') = \sum_{k=-\infty}^\infty
\int_{-\infty}^\infty \frac{d\omega}{2\pi}  
e^{-i [k \Omega_0 t + \omega (t-t') ]} C^<_{\alpha \beta}(k,\omega).
\ee

Substituting with the Floquet-Fourier expansions of the Green's
functions (see Eq. (\ref{floquet})) into Eq. (\ref{C-lesser}) and
rewriting into the form of Eq. (\ref{C-lesser-floquet}) we can obtain
the $k=0$ Floquet-Fourier component, which is
\be \label{lesser-ABCD}
 C^<_{\alpha \beta} (0,\omega) = i \left[
   \mathcal{A}_{\alpha\beta}^<(\omega)
   + \mathcal{A}_{\alpha\beta}^>(-\omega)^*
   + \mathcal{B}_{\alpha\beta}^<(\omega)
   + \mathcal{B}_{\alpha\beta}^>(-\omega)^* \right], 
\ee
where
\ba \label{ABCD}
\mathcal{A}_{\alpha\beta}^<(\omega) & = & \sum_{k=-\infty}^\infty
\int_{-\infty}^\infty 
\frac{d\omega'}{2\pi}  
w_{c\alpha} \sum_{k\alpha} G^>_{l\beta,k\alpha}(-k,\omega'_k) 
\nonumber \\ & &
\times w_{c\beta} \sum_{k\beta} G^<_{l\alpha,k\beta}(k,\omega'+\omega),
\nonumber \\
\mathcal{B}_{\alpha\beta}^<(\omega) & = & \sum_{k=-\infty}^\infty \int_{-\infty}^\infty
\frac{d\omega'}{2\pi}  
G^>_{l\alpha,l\beta}(k,\omega')^* 
\nonumber \\ & &
\times w_{c\alpha} w_{c\beta} \sum_{k\alpha,k\beta}
G^<_{k\alpha,k\beta}(k,\omega'+\omega), 
\nonumber
\ea
and $\mathcal{X}^>$ is obtained from $\mathcal{X}^<$ by switching
$<$ with $>$ ($\mathcal{X} = \mathcal{A}, \mathcal{B}$). 
%where the property $G^{<,>}_{j,j'}(k,\omega) =
%- G^{<,>}_{j',j}(-k,\omega_k)^*$ was used.
Since the greater Floquet-Fourier component is obtained from
Eq. (\ref{C-lesser}) by switching $<$ with $>$, it can be written as:
\be
C^>_{\alpha \beta} (0,\omega) = i \left[
   \mathcal{A}_{\alpha\beta}^>(\omega)
   + \mathcal{A}_{\alpha\beta}^<(-\omega)^*
   + \mathcal{B}_{\alpha\beta}^>(\omega)
   + \mathcal{B}_{\alpha\beta}^<(-\omega)^* \right].
\ee
where $\mathcal{X}^>$ is obtained from $\mathcal{X}^<$ by switching
$<$ with $>$ ($\mathcal{X} = \mathcal{A}, \mathcal{B}, \mathcal{C},
\mathcal{D}$). 

We want to obtain an expression of the previous quantities in terms of
$G^R_{l,l'}(k,\omega)$ with $l,l'$ in the central system. In order to
do so we need to calculate the Floquet-Fourier components of
\ba
\sum_{k\alpha} G^<_{l\beta,k\alpha}(t,t') & = & 
\frac{1}{w_{c\alpha}} \int d t_1  
\left[ G^R_{l\beta,l\alpha}(t,t_1) \Sigma^<_\alpha(t_1,t') \right.
\nonumber \\ &&
\left. + G^<_{l\beta,l\alpha}(t,t_1) \Sigma^A_\alpha(t_1,t') \right],
\ea
and 
\ba
\sum_{k\alpha,k\beta} G^<_{k\alpha,k\beta}(t,t') & = &
\frac{1}{w_{c\alpha} w_{c\beta}} \bigg\{ \delta_{\alpha \beta} \,
\Sigma^<_\alpha(t,t') + \int  dt_1 dt_2
\nonumber \\ &&
\times \Big[ \Sigma^R_\alpha(t,t_1)
G^R_{l\alpha,l\beta}(t_1,t_2) \Sigma^<_\beta(t_2,t')
\nonumber \\ &&
+ \Sigma^R_\alpha(t,t_1) G^<_{l\alpha,l\beta}(t_1,t_2) 
\Sigma^A_\beta(t_2,t')  
\nonumber \\ &&
+ \Sigma^<_\alpha(t,t_1) G^A_{l\alpha,l\beta}(t_1,t_2)
\Sigma^A_\beta(t_2,t') \Big] \bigg\},
\nonumber \\
\ea
where
\ba
\Sigma^<_\alpha(t,t') & = & |w_{c\alpha}|^2 \sum_{k\alpha}
g^{0,<}_{k\alpha,k\alpha} (t,t'), 
\nonumber \\ 
\Sigma^R_\alpha(t,t') & = & |w_{c\alpha}|^2 \sum_{k\alpha}
g^{0,R}_{k\alpha,k\alpha} (t,t'),
\nonumber \\ 
\Sigma^A_\alpha(t,t') & = & \Sigma^R_\alpha(t',t)^*,
\ea
with their respective Fourier transforms
\ba
\Sigma^<_\alpha(\omega) & = & i f_\alpha(\omega) \Gamma_\alpha(\omega), \nn \\
\Sigma^R_\alpha(\omega) & = & \int_{-\infty}^\infty \frac{d\omega'}{2\pi}
\frac{\Gamma_\alpha(\omega')}{\omega - \omega' + i 0^+}, \nn \\
\Sigma^A_\alpha(\omega) & = & \Sigma^R_\alpha(\omega)^*.
\ea

Using the Fourier transforms of $\Sigma^<$ and $\Sigma^R$ together
with the Floquet-Fourier expansion for the Green's functions (see
Eq. (\ref{floquet})) we obtain
\ba \label{lesser-1}
\sum_{k\alpha} G^<_{l\beta,k\alpha}(k,\omega) & = & 
 \frac{1}{w_{c\alpha}} \Big\{ G^R_{l\beta,l\alpha}(k,\omega)
\Sigma^<_\alpha(\omega) 
\nonumber \\ &&
+ G^<_{l\beta,l\alpha}(k,\omega) \Sigma^R_\alpha(\omega)^* \Big\},
\ea
where
\ba \label{Glesser}
G^\lessgtr_{l\beta,l\alpha}(k,\omega') & = & \sum_{m=-\infty}^\infty
\sum_\gamma \Sigma^\lessgtr_\gamma(\omega'_m) 
G^R_{l\alpha,l\gamma}(-m,\omega'_m)^*
\nonumber \\ &&
\times G^R_{l\beta,l\gamma}(k-m,\omega'_m),
\ea
with $\Sigma^\lessgtr_\gamma(\omega) =  \Gamma_\gamma(\omega)
\lambda^\lessgtr_\gamma(\omega)$, and 
\ba \label{lambda}
\lambda^<_\gamma(\omega) & = & i f_\gamma(\omega),
\nonumber \\ 
\lambda^>_\gamma(\omega) & = & -i (1-f_\gamma(\omega)).
\ea

On the other hand, 
\ba \label{lesser-2}
\sum_{k\alpha, k\beta} G^<_{k\alpha,k\beta}(k,\omega) & = & 
\frac{1}{w_{c\alpha}w_{c\beta}} \Big\{ \delta_{\alpha \beta}
\delta_{k 0} \Sigma^<_\alpha(\omega)
\nonumber \\ &&
+ G^R_{l\alpha,l\beta}(k,\omega)
\Sigma^<_\beta(\omega) \Sigma^R_\alpha(\omega_k)
\nonumber \\ &&
+ \Big[ G^R_{l\beta,l\alpha}(-k,\omega_k)^*
\Sigma^<_\alpha(\omega_k) 
\nonumber \\ &&
+ G^<_{l\alpha,l\beta}(k,\omega) \Sigma^R_\alpha(\omega_k) \Big]
\Sigma^R_\beta(\omega)^* \Big\}.
\nonumber \\ 
\ea

By substituting Eqs. (\ref{lesser-1}) and (\ref{lesser-2}) into
Eq. (\ref{ABCD}) and then into Eq. (\ref{lesser-ABCD}) we obtain an
expression for $C^<_{\alpha \beta} (0,\omega)$. The expression for the
greater Green's function is obtained by switching $<$ with $>$.

We are interested in the case where $\alpha = \beta = P$, {\em i.e.}
we are interested in fluctuations in the current flowing to the
probe. As we are dealing with a non-invasive probe, we only keep terms
up to the lowest order in the coupling between the system and the
probe. Since $\Gamma_\alpha, H_\alpha \propto |w_{c\alpha}|^2$ it is
easy to see that

\ba
\mathcal{A}_{PP}^<(\omega) & = & O \left(|w_{cP}|^4 \right),
\nonumber \\
\mathcal{B}_{PP}^<(\omega) & = &  \int_{-\infty}^\infty \frac{d\omega'}{2\pi}
G^>_{lP,lP}(0,\omega')^* 
\lambda_P^<(\omega' + \omega) \Gamma_P(\omega'+\omega)
\nonumber \\ &&
+ O \left(|w_{cP}|^4 \right).
\ea
Thus, $\mathcal{B}_{PP}^<(\omega)$ and $\mathcal{B}_{PP}^>(\omega)$
are the only terms with contributions of order $|w_{cP}|^2$. Using
Eqs. (\ref{Glesser}) and (\ref{lambda}) we can write, up to this
order, 
\ba
C^K_{PP}(0,\omega) & = & -i \int_{-\infty}^\infty
\frac{d\omega'}{2\pi} \sum_{k=-\infty}^\infty 
\sum_{\gamma=L,R} |G^R_{lP,l\gamma}(-n,\omega'_n)|^2
\nonumber \\ &&
\times \Big\{ f_{\gamma P}(\omega'_n,\omega'+\omega)
\Gamma_P(\omega' + \omega) 
\nonumber \\ &&
+ f_{\gamma P}(\omega'_n,\omega'-\omega) 
\Gamma_P(\omega' - \omega) \Big\} \Gamma_\gamma(\omega'_n) .
\nonumber \\
\ea

To obtain an expression for $\overline{\varphi}_P^*(\omega)$ we
can use the following identity, easily deduced from
Eq. (\ref{greenFF}), 
\be
\overline{\varphi}_P^*(\omega) \equiv - 2 \mbox{Im}
         [C^R_{PP}(0,\omega)] = i \left( C^>_{PP}(0,\omega) -
         C^<_{PP}(0,\omega) \right), 
\ee
which leads to
\ba
\overline{\varphi}_P^*(\omega) & = & \int_{-\infty}^\infty
\frac{d\omega'}{2\pi} \sum_{k=-\infty}^\infty \sum_{\gamma=L,R}
\Gamma_\gamma(\omega'_n) |G^R_{lP,l\gamma}(-n,\omega'_n)|^2 
\nonumber \\ &&
\times \Big\{ \Big[f_\gamma(\omega'_n) - f_P(\omega'+\omega) \Big]
\Gamma_P(\omega' + \omega) 
\nonumber \\ &&
- \Big[f_\gamma(\omega'_n) - f_P(\omega'-\omega) \Big]
\Gamma_P(\omega' - \omega) \Big\} .
\ea

If we additionally consider a constant density of states in the probe
to obtain a result independent of any particular probe, we arrive to
the expressions given in Eqs. (\ref{ccK}) and (\ref{ccR}).

%\newpage

\end{document}